# Bipolar Charge Plasma Transistor: A Novel Three Terminal Device

M. Jagadesh Kumar *Senior Member, IEEE* and Kanika Nadda,

*Abstract*— A distinctive approach for forming a lateral Bipolar Charge Plasma Transistor (BCPT) is explored using 2-D simulations. Different metal work-function electrodes are used to induce n- and p-type charge plasma layers on undoped SOI to form the emitter, base and collector regions of a lateral NPN transistor. Electrical characteristics of the proposed device are simulated and compared with that of a conventionally doped lateral bipolar junction transistor with identical dimensions. Our simulation results demonstrate that the BCPT concept will help us realize a superior bipolar transistor in terms of a high current gain compared to a conventional BJT. This BCPT concept is suitable in overcoming doping issues such as dopant activation and high-thermal budgets which are serious issues in ultra thin SOI structures.

## I. INTRODUCTION

BIPOLAR transistors exhibit a number of significant advantages such as well controllable characteristics, high speed, high gain and low output resistance. These are excellent properties for mixed signal circuit design and analog amplifiers. An emergent trend in modern high density VLSI circuits is the integration of bipolar transistors with CMOS technology on thin SOI films.

However, CMOS fabrication favors lateral bipolar transistor concepts [1-6] on SOI since it is easier to tune the SOI layer properties to optimize the CMOS devices without degrading the performance of the bipolar devices [2]. Realization of a true low cost complementary-BiCMOS process is possible only when the lateral bipolar transistors can be fabricated on CMOS using simple fabrication steps with low thermal budgets. High temperature cycles, required for post ion-implantation annealing of emitter and base regions of a bipolar transistor can create complications while integrating the bipolar process with the CMOS process.

To overcome the above problems, we propose for the first time a new device concept designated as Bipolar Charge Plasma Transistor (BCPT) as shown in Fig. 1. In this structure, no ion implantation is done on silicon; rather, the p-type and n-type regions are formed in the undoped silicon layer by employing metal electrodes with different work-functions [7-9] to induce (i) an electron plasma for forming the emitter and collector regions and (ii) hole plasma to form the base region.

The most important advantage of the proposed BCPT is the absence of doped regions. This avoids the necessity of complicated thermal budgets using expensive annealing equipment.

Using two dimensional simulations, we describe in this paper, the unique electrical characteristics of the BCPT on an undoped SOI. Our results show that the BCPT exhibits a good transistor action with a significantly larger current gain compared to a conventional BJT with the same device geometry. Our results may provide the incentive for further experimental exploration of the BCPT concept since a charge-plasma pn junction has already been demonstrated experimentally [7].

## II. DEVICE STRUCTURE AND SIMULATION PARAMETERS

The schematic views of the lateral NPN BCPT and the conventional lateral BJT are shown in Fig. 1. The parameters of the devices used in our simulation are: background doping (N) $=1\times10^{13}$ /cm$^3$, buried-oxide layer thickness ($t_{BOX}$) = 375 nm, separation between the electrodes ($L_S$) = 100 nm, gate oxide layer thickness ($t_{ox}$) = 5 nm and undoped silicon layer thickness ($t_{Si}$) = 15 nm. In BCPT, the electrode length of the emitter, base and collector regions is chosen to be 0.2 μm, 0.1 μm and 0.4 μm, respectively. For creating the emitter, by inducing electrons in the undoped silicon body, Hafnium (work-function $\varphi_{m,E}$ =3.9 eV) is employed as the emitter electrode metal. For inducing holes to create the base region, platinum (work function $\varphi_{m,B}$= 5.65 eV) is used. As the collector of an npn transistor needs to have a lower carrier concentration than the base region, to induce electrons to create the collector region, Al (work-function $\varphi_{m,C}$ of 4.28 eV) is used. It is also important to choose an appropriate thickness for the silicon film. To maintain uniform induced carrier distribution throughout the silicon thickness, from the oxide-Si interface to the Si-buried oxide interface, the silicon film thickness has to be kept within the Debye length, i.e.,

$$L_D = \sqrt{(\varepsilon_{Si} \cdot v_T)/(q \cdot N)}$$

where $\varepsilon_{Si}$ is the dielectric constant of silicon, $V_T$ is the thermal voltage, and N is the carrier concentration in the body [8]. For the BCPT structure shown in Fig. 1, the simulated net carrier concentrations (taken at 2 nm away from the Si-oxide interface) for zero bias and typical forward active conditions are shown in Fig. 2. Either under thermal equilibrium condition ($V_{BE}$ = 0 V and $V_{CE}$ = 0 V) or under forward active bias condition ($V_{BE}$ = 0.7 V and $V_{CE}$ = 1 V), the electron and hole charge plasma concentration is maintained in the emitter ($n_E$ = 2×10$^{18}$ /cm$^3$), base ($p_B$ = 1×10$^{20}$ /cm$^3$) and collector ($n_C$ = 2×10$^{17}$ /cm$^3$) regions due to the choice of work functions of the metal electrodes with respect to that of silicon e.g., low value of $\varphi_{m,E}$ and $\varphi_{m,C}$, for the emitter and collector electrodes and higher

Manuscript received November 17, 2011. (Write the date on which you submitted your paper for review.)
The authors are with the Department of Electrical Engineering, Indian Institute of Technology, New Delhi 110 016, India (e-mail: mamidala@ee.iitd.ac.in, kanika.shippu@gmail.com ).

$\varphi_{m,B}$ for the base electrode. This presence of charge plasma (electron/hole) concentrations in the undoped silicon film permits the formation of a bipolar charge plasma transistor.

Quasi-Fermi levels (QFL) of the BCPT under thermal equilibrium and under forward active bias conditions are shown in Fig. 3. Under thermal equilibrium, the quasi-Fermi levels for electrons and holes align with each other as in the conventional BJTs as shown in Fig. 3(a). Since excess carriers are present on either side of the forward biased emitter-base junction, the quasi-Fermi levels are different for holes and electrons in this case. In the reverse biased base-collector region, the minority carrier concentration in the collector region does not reach its thermal equilibrium value because of the presence of the field electrode. This is why the quasi-Fermi levels do not merge throughout the collector region as shown in Fig. 3 (b).

The current in the BCPT is strongly controlled by the electrodes. The lateral NPN transistor with which we have compared our results is also an SOI structure and has the same device parameters as that of the BCPT except that the emitter, base and the collector regions have a length of 0.25 μm, 0.2 μm and 0.45 μm, and a typical doping of $N_D = 1\times10^{20}$ /cm$^3$, $N_A = 9\times10^{17}$ /cm$^3$ and $N_D = 2\times10^{17}$ /cm$^3$, respectively. The lengths are chosen so as to have an equal neutral base width for both the transistors. The doping profile is chosen similar to that of a practical bipolar transistor. We could not choose the emitter and base doping of the BJT same as the charge plasma concentrations in the BCPT since this would make the current gain of the BJT extremely low (less than 1) due to the larger base Gummel number compared to the emitter Gummel number in BCPT.

Simulations have been performed with ATLAS device simulation tool [10] using the Fermi Dirac distribution for carrier statistics, with Philip's unified mobility model [11], and doping-induced band-gap narrowing model [12], all with default silicon parameters. Standard thermionic emission model [9] is invoked for the emitter contact of BCPT with a work function of 3.9 eV and a surface recombination velocity of $2.2\times10^6$ cm/s and $1.6\times10^6$ cm/s for electrons and holes, respectively. The same base metal i.e. platinum is used at the base contact of BCPT and BJT to ensure that the base contact properties are identical in both the transistors. This ensures that the lower base current in the BCPT is not due to a difference in the base contact properties. Ohmic contact conditions are assumed at all other contacts. The metal-semiconductor contact resistances are assumed to be negligible in both the transistors. For recombination, we have enabled Klassen's model for concentration dependent lifetimes for SRH recombination with intrinsic carrier lifetimes $n_{ie} = n_{ih} = 0.2$ μs [13]. High electric field velocity saturation is modeled through the field dependent mobility model [10]. The screening effects in the inversion layer are also considered by invoking the Shirahata mobility model [14]. Selberherr impact ionization model is used for calculating $B_{VCEO}$ [15]. To ensure that our simulations are accurate, we have first reproduced the simulation results of the charge-plasma diode characteristics [8] using the above models and ensured that the current voltage characteristics predicted by our simulation match with those reported in [8].

## III. RESULTS AND DISCUSSION

The gummel plot for the BCPT is compared with that of the BJT as shown in Fig. 4. While the BCPT exhibits almost the same collector current as compared to the conventionally doped BJT, the base current of the BCPT is much lower than that of the BJT due to the SALTRAN effect [5, 16]. When a lightly doped emitter is contacted with a metal whose work function is less than that of silicon, it results in a surface accumulation of electrons at the metal contact as shown in Fig. 5. However, no such electron accumulation is observed in the case of the BJT since its emitter is heavily doped. In the case of BCPT, the accumulation of electrons results in an electric field due to the electron concentration gradient from the metal–semiconductor interface toward the emitter–base junction. The direction of this field is such that it repels the minority holes injected from the base into the emitter, resulting in a reduced hole concentration gradient in the emitter region. As a consequence, there is a reduction in base current, leading to a significant improvement in the current gain as explained in [5, 16]. Hence, the peak current gain of the BCPT (4532) is several orders higher compared to the peak current gain of the conventional BJT (26.5) as shown in Fig. 6. As discussed in [8], [17] the metal work-functions strongly control the current in the charge plasma diode. Likewise the work-functions of the emitter and base electrode metals strongly control the emitter and collector current. By appropriately choosing the metal electrode work-functions, the BCPT characteristics can be effectively controlled. The output current drive of the BCPT is in agreement with the forward current levels of the charge plasma diode reported in [7] and [8]. However, further work needs to be done to increase the current levels by optimizing the device dimensions and the choice of metals.

The output characteristics of the BCPT are compared with that of the conventional BJT in Fig. 7. Due to its larger current gain, the BCPT exhibits a slightly lower collector breakdown voltage compared to the conventional BJT. However, we observe negligible base width modulation in the BCPT due to a higher concentration of induced holes in the base region of the BCPT.

The saturation voltage (VCE (sat)) is marginally higher for BCPT due to the lower electron concentration in the emitter compared to the BJT. If the emitter doping in the BJT structure is made similar to the electron concentration in the emitter of BCPT, we have observed a similar increase in VCE (sat) even for the BJT. For bipolar transistors to be useful in high-frequency applications, their $f_T$ should be higher than 1 GHz. Fig. 8 shows that the simulated cut-off frequency $f_T$ of the BCPT (3.7 GHz) and the BJT structure (7.5 GHz). However, the presence of intrinsic gaps Ls (100 nm) between the emitter-base and collector base junctions which are larger than in the BJT will lower the cut-off frequency of the BCPT compared to that of the BJT structure. Our simulation results show that the speed of the BCPT is lower than that of the conventional lateral BJTs and further work needs to be done to optimize the speed of BCPT.

### IV. EFFECT OF INTERFACE TRAPS ON CURRENT GAIN

A key feature of the BCPT structure is the low work-function metal electrode employed at the emitter contact forming a metal-semiconductor junction (hafnium-silicon). Depending on the surface preparation and metal deposition methods, there is a definite possibility of the presence of traps at the hafnium-silicon interface due to lattice discontinuity and the dangling bonds, not to mention inter-diffusion. These traps are very likely to affect the electron distribution and the electric field in this area, which in turn would affect the SALTran effect [18]. Both the acceptor and the donor type of interface traps can affect the accumulation of electrons at the interface and thus the current gain of the BCPT.

To simulate the influence of interface traps on the device characteristics, we have considered a typical range of interface traps which may be present at the interface. In our simulations both the acceptor and the donor type of traps are introduced with the trap energy level (E.level) at 0.49 eV from the conduction (or valance) band [5]. The degeneracy factor (degen) for the trap level is chosen to be 12 and the capture cross-sections for electrons (sign) and holes (sigp) are taken to be $2.85 \times 10^{-15}$ cm$^{-2}$ and $2.85 \times 10^{-14}$ cm$^{-2}$, respectively.

Peak current gain reduces when the trap density of the acceptor (or donor) type of interface traps increases at the interface as shown in Fig. 9 and the effect is even more severe when both type of traps are present at the interface. This effect is also observed for a conventional SALTran BJT in [5]. However, even with a trap density of $1 \times 10^{11}$ cm$^{-2}$, BCPT still maintains a substantially high peak current gain as observed in Fig. 9.

### V. CONCLUSION

In conclusion, we report the possibility of realizing a unique Bipolar Charge Plasma Transistor (BCPT) made of electron and hole plasma on undoped silicon making it immune to thermal budget concerns faced by doped transistors and CMOS devices. The efficacy of the proposed concept is verified using two-dimensional numerical simulations. Our simulation results demonstrate that the BCPT exhibits a significantly large current gain compared to a conventional BJT with doped regions of similar geometry. The BCPT concept can also be applied for materials such as SiC in which doping is an issue or for hetero-junction bipolar transistors. This idea can be extended to an undoped silicon nanowire with surround gate electrodes to induce the emitter, base and collector regions making it compatible with the future nanowire and FinFET based CMOS technology [19]. Since the charge plasma based diode concept has already been demonstrated experimentally [7], we believe that our results will provide the incentive for further experimental verification of the BCPT concept. However, it remains to be seen whether the fabrication of BCPT will be cost effective, since three different metals are required for the emitter, base and collector electrodes. The device also needs to be optimized to match the speed performance of the commercially available lateral bipolar transistor technology.

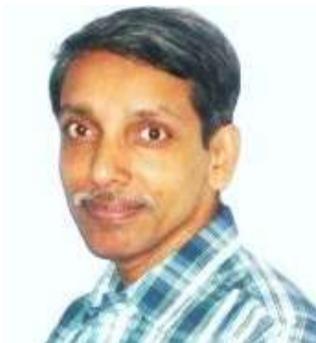 **M. Jagadesh Kumar** was born in Mamidala, Andhra Pradesh, India. He received the M.S. and Ph.D. degrees in electrical engineering from the Indian Institute of Technology (IIT), Madras, India.

From 1991 to 1994, he performed postdoctoral research on the modelling and processing of high-speed bipolar transistors at the Department of Electrical and Computer Engineering, University of Waterloo, Waterloo, ON, Canada. While with the University of Waterloo, he also did research on amorphous-silicon thin-film transistors. From July 1994 to December 1995, he was initially with the Department of Electronics and Electrical Communication Engineering, IIT, Kharagpur, India, and then, he was with the Department of Electrical Engineering, IIT, New Delhi, India, where he became an Associate Professor in July 1997 and has been a Full Professor in January 2005. He is currently the Chair Professor of the NXP (Philips) (currently, NXP Semiconductors India Pvt. Ltd.) established at IIT Delhi by Philips Semiconductors, The Netherlands. He is the Coordinator of the Very Large Scale Integration (VLSI) Design, Tools, and Technology interdisciplinary program at IIT Delhi. He is also a Principal Investigator of the Nano-scale Research Facility at IIT Delhi.

His research interests include nanoelectronic devices, device modelling and simulation for nanoscale applications, integrated-circuit technology, and power semiconductor devices. He has published extensively in these areas of research with four book chapters and more than 150 publications in refereed journals and conferences. His teaching has often been rated as outstanding by the Faculty Appraisal Committee, IIT Delhi.

Dr. Kumar is a fellow of the Indian National Academy of Engineering, The National Academy of Sciences, India and the Institution of Electronics and Telecommunication Engineers (IETE), India. He is recognized as a Distinguished Lecturer of the IEEE Electron Devices Society (EDS). He is a member of the EDS Publications Committee and the EDS Educational Activities Committee. He is an Editor of the IEEE TRANSACTIONS ON ELECTRON DEVICES and an Associate Editor of IEEE Technology News. He was the lead Guest Editor for the following: 1) the joint special issue of the IEEE TRANSACTIONS ON ELECTRON DEVICES and the IEEE TRANSACTIONS ON NANOTECHNOLOGY (November 2008 issue) on Nanowire Transistors: Modelling, Device Design, and Technology and 2) the special issue of the IEEE TRANSACTIONS ON ELECTRON DEVICES on Light Emitting Diodes (January 2010 issue). He is the Editor-in-Chief of the IETE Technical Review and an Associate Editor of the Journal of Computational Electronics. He is on the editorial board of Recent Patents on Nanotechnology, Recent Patents on Electrical Engineering, Journal of Low Power Electronics, and Journal of Nanoscience and Nanotechnology. He has reviewed extensively for different international journals.

He was a recipient of the 29th IETE Ram Lal Wadhwa Gold Medal for his distinguished contribution in the field of semiconductor device design and modelling. He was also the first recipient of the India Semiconductor Association–VLSI Society of India TechnoMentor Award given by the India Semiconductor Association to recognize a distinguished Indian academician for playing a significant role as a Mentor and Researcher. He is also a recipient of the 2008 IBM Faculty Award. He was the Chairman of the Fellowship Committee of The Sixteenth International Conference on VLSI Design (January 4–8, 2003, New Delhi, India), the Chairman of the Technical Committee for High Frequency Devices of the International Workshop on the Physics of Semiconductor Devices (December 13–17, 2005, New Delhi), the Student Track Chairman of the 22nd International Conference on VLSI Design (January 5–9, 2009, New Delhi), and the Program Committee Chairman of the Second International Workshop on Electron Devices and Semiconductor Technology (June 1–2, 2009, Mumbai, India).
For more details, please visit http://web.iitd.ac.in/~mamidala

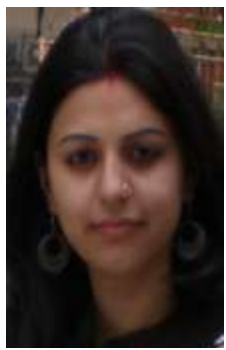 **Kanika Nadda** received the B-Tech degree in electronics and communication from HP University, H.P., India, in 2006. She received the M-Tech degree in VLSI design and automation form NIT Hamirpur, H.P., India. She is currently working toward the PhD degree in the Department of Electrical Engineering, Indian Institute of Technology, Delhi, India. Her research interests include nano-scale bipolar devices and modeling

.

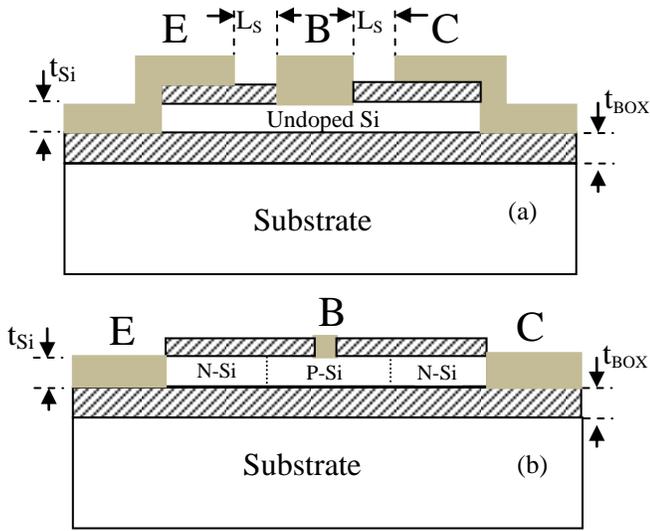

Fig. 1. Schematic cross-sectional view of (a) the BCPT and (b) the conventional BJT.

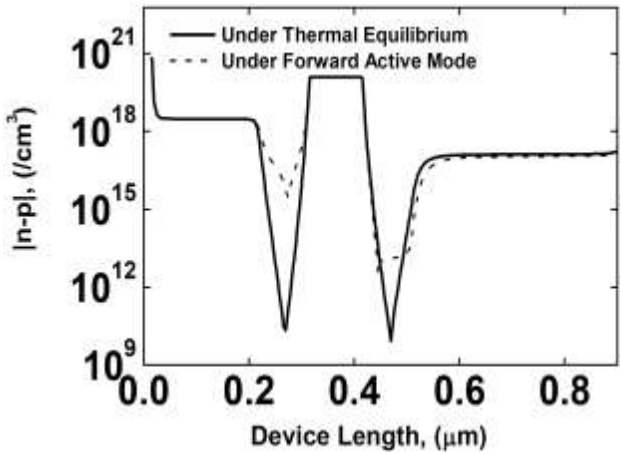

Fig. 2. Simulated net carrier concentrations in the BCPT for different bias conditions.

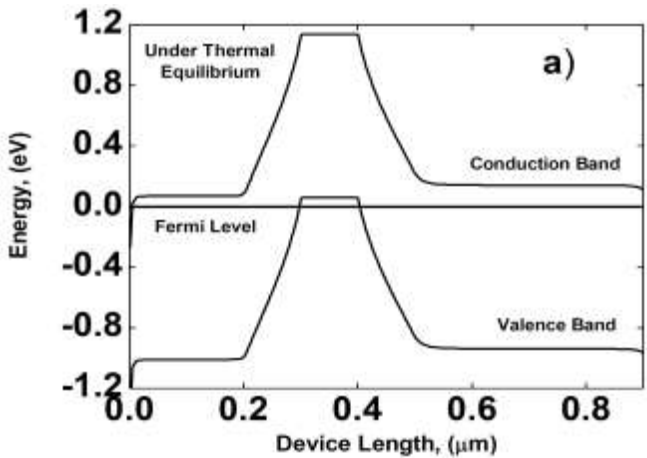

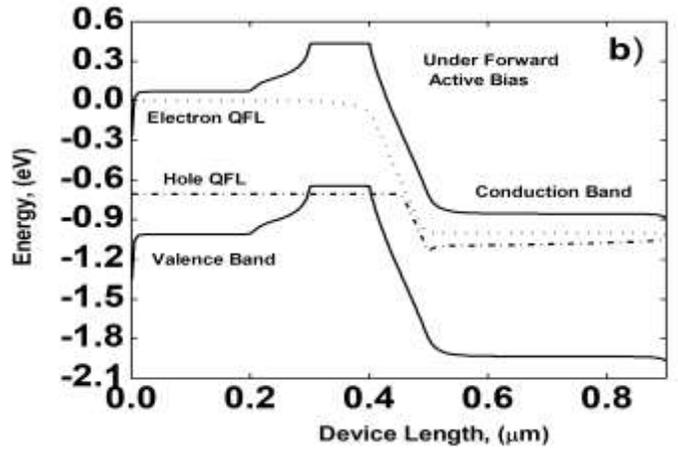

Fig. 3. Simulated energy band diagram (a) under thermal equilibrium and (b) under forward active bias conditions for the BCPT.

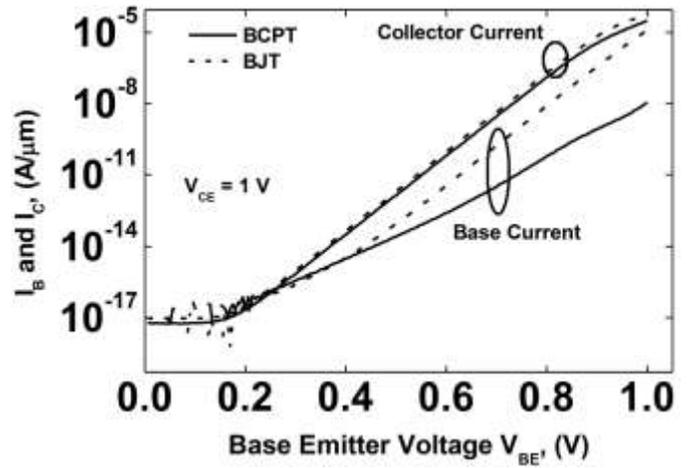

Fig. 4. Gummel plots of the BCPT and the conventional BJT structures.

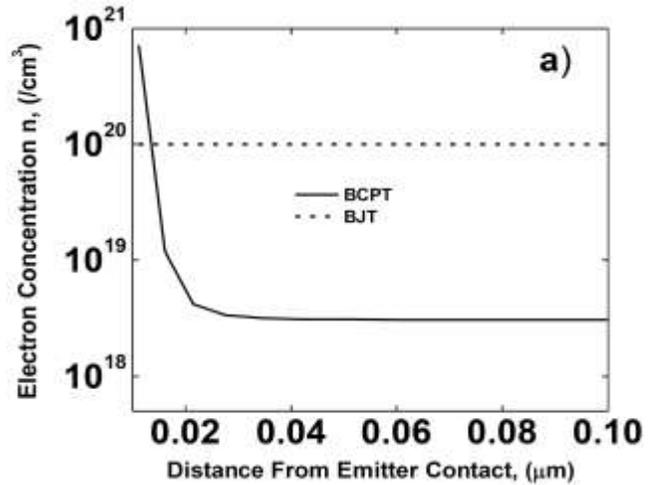

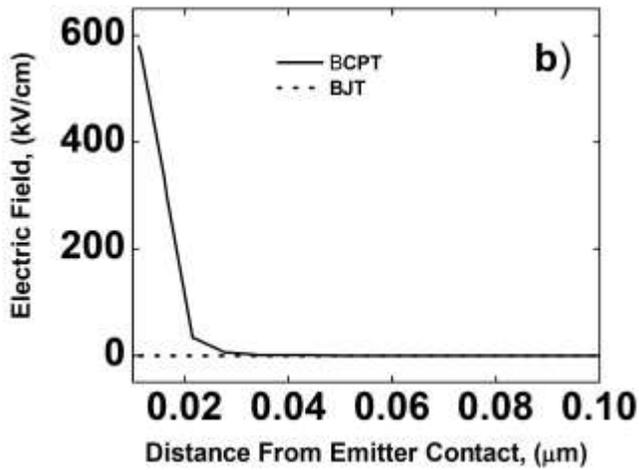

Fig. 5. (a) Electron Concentration and (b) Electric field distribution in the emitter region of the BCPT and the conventional BJT structures.

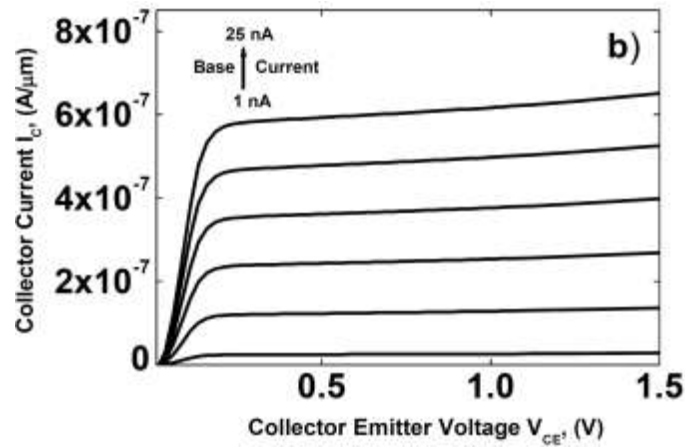

Fig. 7. Output characteristics of (a) the BCPT and (b) the conventional BJT.

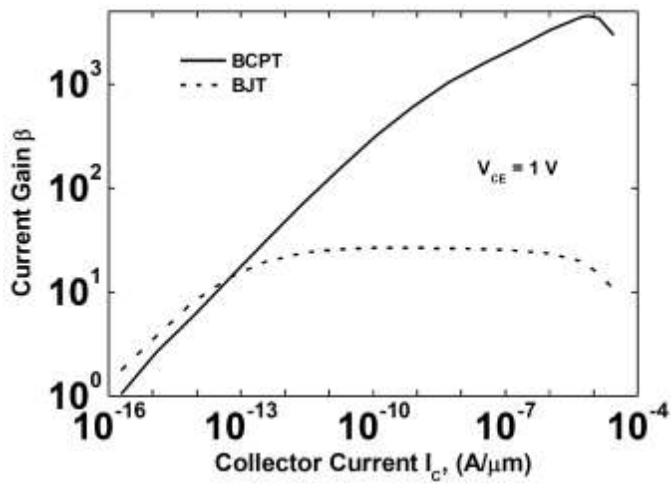

Fig. 6 Current gain variation of the BCPT and the conventional BJT.

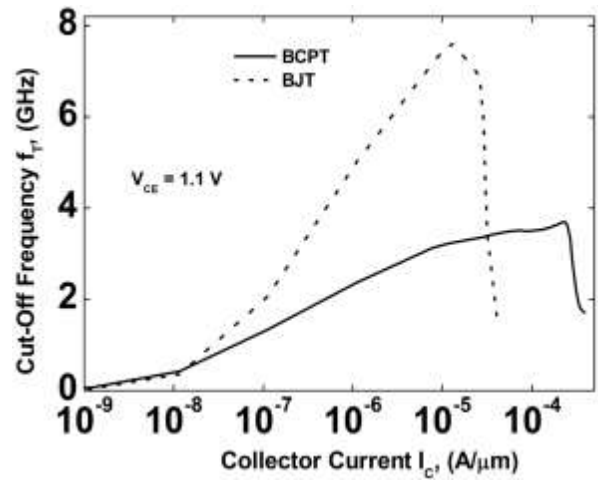

Fig. 8. Cut-Off frequency of (a) the BCPT and (b) the conventional BJT.

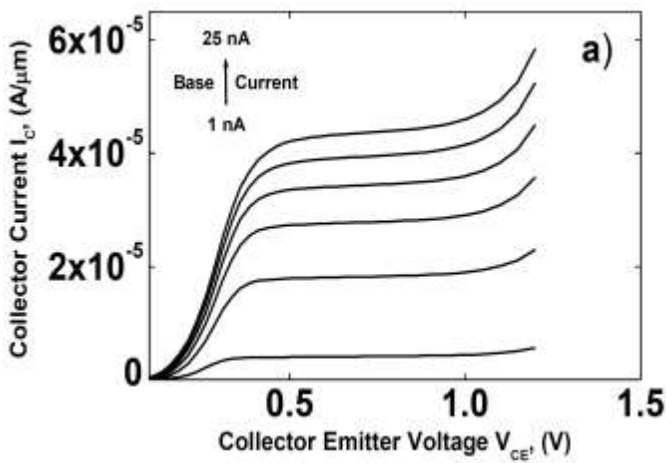

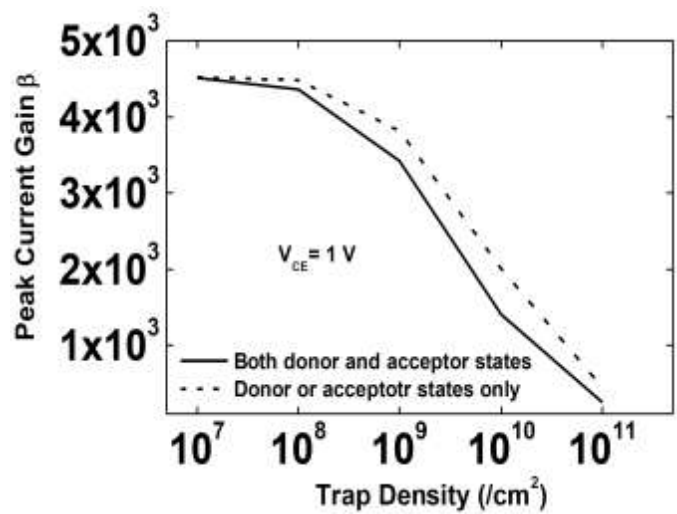

Fig. 9. Peak current gain versus trap density for the BCPT structure.